\documentclass[pra,twocolumn,showpacs,floatfix]{revtex4}

\usepackage{amssymb,amsmath,amsthm}
\usepackage{graphicx}
\usepackage[all]{xy}

\newtheorem{Theorem}{Theorem}

\newcommand*{\ii}[0]{{\it i}}
\newcommand*{\eps}[0]{{\epsilon}}

\newcommand*{\C}[0]{{\mathbb C}}

\begin{document}

\title{Optimal quantum control in nanostructures: Theory and application to generic three-level system}

\author{Alfio~Borz\`{\i}}
\author{Georg Stadler}
\affiliation{Institut f\"ur Mathematik, Karl-Franzens-Universit\"at Graz, 
Heinrichstra\ss e 36, 8010 Graz, Austria}

\author{Ulrich Hohenester}\email{ulrich.hohenester@uni-graz.at} 
\affiliation{Institut f\"ur Theoretische Physik, 
Karl-Franzens-Universit\"at Graz, Universit\"atsplatz 5, 8010 Graz, Austria}


%

\begin{abstract}

Coherent carrier control in quantum nanostructures is studied within the framework of {\em optimal control}.\/ We develop a general solution scheme for the optimization of an external control (e.g., lasers pulses), which allows to channel the system's wavefunction between two given states in its most efficient way; physically motivated constraints, such as limited laser resources or population suppression of certain states, can be accounted for through a general cost functional. Using a generic three-level scheme for the quantum system, we demonstrate the applicability of our approach and identify the pertinent calculation and convergence parameters.
\end{abstract}


\maketitle

\section{Introduction}

Recent years have witnessed enormous interest in controlling quantum phenomena in a variety of nanoscale systems \cite{poetz:99}. Quite generally, such control allows to modify the system's wavefunction at will through appropriate tailoring of external fields, e.g., laser pulses: while in {\em quantum optics}\/ the primary interest of this wavefunction engineering lies on the exploitation of quantum coherence among a few atomic levels \cite{scully:97}, in {\em quantum chemistry}\/ optical control of molecular states has even led to the demonstration of optically driven chemical reactions of complex molecules \cite{rabitz:00}; furthermore, starting with the seminal work of Heberle {\em et al.}\/ \cite{heberle:95} coherent-carrier control in semiconductors and semiconductor nanostructures has recently been established as a mature field of research on its own. In particular with the advent of semiconductor quantum dots \cite{hawrylak:98}, sometimes referred to as {\em artificial atoms},\/ one now has a system at hand which resembles many of the atomic properties whilst offering at the same time all the flexibility of semiconductor nanostructures: experimentally basic quantum-coherence phenomena such as polarization beating \cite{bonadeo:98b} or Rabi-type flopping \cite{stievater:01} have been demonstrated, whereas theoretically effects such as coherent population transfer \cite{hohenester.apl:00,pazy:01} or entanglement control \cite{biolatti:00,chen:01b,piermarocchi:02} have been proposed.

In the last few years this research area has received further impetus from the emerging fields of quantum computation and quantum communication \cite{bouwmeester:00}, aiming at quantum devices where the wavefunction can be manipulated with highest possible precision ({\em quantum gates}). This high-fidelity quantum-state engineering calls for strategies which allow an optimal suppression of environment losses during gating; self-evidently, such outstanding performance can only be achieved if the most sophisticated experimental and theoretical techniques for optical control of quantum nanostructures are put together. It is worth emphasizing that hitherto there exists no clear consensus of how to optimally tailor the system's control, and it appears that each field of research has come up with its own strategies: for instance, quantum-optical implementations in atoms benefit from the long atomic coherence times of meta-stable states, and it usually suffices to rely on effective models which can be grasped from the solution of simplified level schemes (e.g., adiabatic population transfer in an effective three-level system \cite{bergmann:98}); in contrast, in quantum chemistry the complexity of molecular states usually does not permit schemes which are solely backed from the underlying level schemes, and learning algorithms, which receive direct feedback from experiment, appear to be the method of choice. Finally, coherent control in semiconductor nanostructures has hitherto been primarily inspired by quantum-optical techniques; however, it is clear that control in future quantum devices will require more sophisticated techniques to account for the enhanced dephasing in the solid states; a first step in this direction has been undertaken in Refs.~\cite{chen:01b,piermarocchi:02}, where the authors have adopted control techniques developed in nuclear-magnetic resonance \cite{ernst:87} to semiconductor nanostructures.

In this paper we examine the problem of coherent-carrier control in quantum nanostructures within the framework of {\em optimal control}\/ \cite{lions,neti,pdr,Fat}. Here, one starts by defining the {\em optimality criteria}\/ (the cost functional); in general, for a desired quantum-state transition this functional will depend on the final state, the wish to suppress population of certain states during the control process, as well as other physically motivated constraints, e.g., limited laser resources. The grand strategy then is to minimize this cost functional and to find the optimal time dependency of the control fields, which, in turn, govern the evolution of quantum states through the underlying dynamic equations (i.e., Schr\"odinger or master equation). The calculation of the necessary optimality conditions for this optimization problem results in a system of coupled equations, which, for high-dimensional systems, may involve heavy computations. Yet, the clear-cut advantage of this optimization approach is the flexibility to steer the control strategies through modification of the cost functional, thus rendering this technique ideally for the purpose of quantum-state engineering.

We have organized our paper as follows. Our theoretical approach is presented in Sec.~\ref{sec:theory}; in Sec.~\ref{sec:algorithm} we derive the numerical algorithm for the solution of the relevant equations. As a first example, in Sec.~\ref{sec:model} we study the optimal control of a generic three-level system. Sec.~\ref{sec:results} summarizes our numerical results, and we finally draw some conclusions in Sec.~\ref{sec:conclusions}.

\section{Theory}\label{sec:theory}

Consider the Schr\"odinger equation for a $n$-component wave 
function $\psi \in L^2(\C^n,[0,T])$:

\begin{equation}
\label{schroe1}
\ii \dot \psi = H \psi , \quad \psi(0) = \psi_0 ,
\end{equation}

\noindent where the Hamiltonian $H=H_0+H_\eps$ accounts for: $H_0$, the unperturbed system; and $H_\eps$, the coupling to an external control field $\eps$, where
 
\begin{equation}\label{est H_eps}
||H_{\eps(t)}||\le K ||\eps(t)||, \:\: K>0
\end{equation}

\noindent is supposed to hold for all $t\in [0,T]$; finally, $\psi_0$ is the initial state of the system ($\hbar=1$ throughout). Note that, strictly speaking, the wavefunction description of Eq.~(\ref{schroe1}) is only allowed for an isolated quantum system. For the problem of our present concern (control in presence of dephasing and relaxation) a more general {\em density-matrix description}\/ would be required \cite{scully:97} to account for the incoherent environment couplings. However, following the procedure outlined in Ref.~\cite{hohenester:02} we observe that even in presence of such coupling it is possible to define a non-Hermitian Hamiltonian of the form (\ref{schroe1}), accounting for dephasing and generalized out-scatterings, if, at the same time, one introduces a further term which accounts for generalized in-scatterings. Thus, since we are aiming at an optimal control of the {\em coherent}\/ time evolution, i.e., we are seeking for solutions which minimize environment losses (see also Sec.~\ref{sec:model}), we can safely neglect in-scattering terms, and we are led to Eq.~(\ref{schroe1}), with $H_0$ being non-Hermitian.

In the following we shall consider the problem of determining the control fields, $\eps \in L^2(\C,[0,T])$, such that Eq.~(\ref{schroe1}) is fulfilled. In so doing we shall be guided by a number of further constraints, which, all together, form the so-called {\em optimal criteria}:\/ firstly, we assert that the control sequence brings the system at time $T$ to the desired state $\psi_d \in \C^n$; secondly, we account for the limited laser resources through a minimization of the control field strengths; thirdly, we may wish to suppress population of intermediate states which suffer strong environment losses (see discussion below). More specifically, all these constraints are summarized in the cost functional:

\begin{eqnarray}
J(\psi, \eps ) &:=& \frac{1}{2} | \psi(T) - \psi_d |_{\C^n}^2
+ \frac{\gamma}{2} || \eps ||_{L^2(\C,[0,T])}^2 \nonumber\\
&&+\frac{1}{2} \sum_{j=1}^n \alpha_j || \psi_j ||_{L^2(\C,[0,T])}^2 .
\label{minJ}
\end{eqnarray}

\noindent where the constants $\gamma>0, \alpha_i\ge 0$ are the weighting factors, which allow to vary the relative importance of the various terms and $\psi_j\in L^2(\C,[0,T])$ denotes the $j$-th component of $\psi$; the last term of (\ref{minJ}) penalizes the occupation of certain components $\psi_j$ during the control process. Apparently, further constraints could be added in a completely similar fashion.
The optimal control problem under consideration can now shortly be written as 
\begin{equation}\label{optcon}
\min J(\psi,\eps) \,\mbox{ subject to \eqref{schroe1}}.
\end{equation}

\noindent We now state that Eq.~\eqref{optcon}  has a solution.

\begin{Theorem}
The optimal control problem \eqref{optcon} admits a solution 
$(\bar{\psi},\bar{\eps})\in H^1(\C^n,[0,T])\times L^2(\C, [0,T])$.

\end{Theorem}

\begin{proof}
The above theorem can be verified in a completely analogous fashion to Ref.~\cite{pdr} (for details see Appendix A and Ref. \cite{Fat}).
\end{proof}

To calculate the necessary optimality conditions of first order for \eqref{optcon}, we use the method of Lagrange multipliers 
\cite{lions} to turn the constrained minimization problem \eqref{optcon} 
into an unconstrained one. 
For this purpose we define the Lagrangian function
\begin{displaymath}
L(\psi,p,\eps)=J(\psi, \eps ) + \Re e \left< p, \ii \dot \psi -  (H_0+H_\eps)\, \psi \right> .
\end{displaymath}
Here, $\left<\phi,\psi\right> = \int_0^T \phi \cdot \psi^* dt$, where `*' means complex 
conjugate and the dot `$\cdot$' denotes the usual vector-scalar product in $\C^n$.

Consider the minimization problem: Find $\tilde \psi$, $\tilde p$, and $\tilde \eps$ such that
$$
L(\tilde \psi,\tilde p,\tilde \eps)
= \inf_{\psi \in X_t^0, \, p \in X_t, \, \eps \in X_t}  L(\psi,p,\eps) ,
$$
where $X_t = L^2(\C^n,[0,T])$ and $X_t^0 = X_t \cap \{\psi: \psi(0) = \psi_0 \}$. 
Here, necessary conditions for a minimum are obtained by equating to zero the Fr\'echet 
derivatives of $L$ with respect to the triple $(\psi, \, p, \, \eps )$. The following 
optimality system is obtained
\begin{subequations}
\begin{eqnarray}
\ii \dot \psi & = &  (H_0+H_\eps)\, \psi , \, \mbox{ with } \psi(0) = \psi_0 , \label{staeq}  \\
\ii \dot p    & = &  (H_0^* + H_\eps) \, p  - q , 
\, \mbox{ with } \ii p(T) = \psi(T) - \psi_d , \label{adjeq} \nonumber\\&& \\
 \eps & = & \frac{1}{\gamma} \Re e [ p \cdot ( \frac{\partial H}{\partial \eps_r} \psi)^*]
   + \ii \, \frac{1}{\gamma} \Re e [ p \cdot ( \frac{\partial H}{\partial \eps_i}\psi)^*] ,  
\label{ctreq}
\end{eqnarray}
\end{subequations}
where $q_j=\alpha_j\psi_j$ and $\eps = \eps_r + \ii \eps_i$. 
Notice that while the state equation (\ref{staeq}) evolves forward in time, 
the adjoint equation (\ref{adjeq}) is marching backwards. The control 
equation (\ref{ctreq}) provides the control function.

\section{Numerical algorithm}\label{sec:algorithm}
In this section we formulate a numerical algorithm that solves the optimality system (\ref{staeq}--c) for given initial and final configurations $\psi_0$ and $\psi_d$, respectively. 
To solve this problem, we apply a gradient-type minimization algorithm, which first determines a search direction with respect to the variable $\eps$ for both the real and imaginary part.  
Then, a simple step-size procedure is applied that guarantees a decrease in the cost functional. 
For given $\eps$ the search direction is calculated as follows: 
the initial condition for the state equation is given by 
$\psi_0$. Once the wave function at $t=T$ is 
computed, the final condition for the adjoint equation is given by $\ii p(T) = \psi(T) - \psi_d$.
Thus, the adjoint variable $p$ can be calculated and the gradient of $J$ with respect to $\epsilon$ can be computed.

Assume that the interval $[0,T]$ has been discretized into a finite number $N_{steps}$ of 
subintervals of size $\delta t$, and $t_m = (m-1)\, \delta t$. A discrete state variable at $t_m$ is 
denoted by $\psi^m$.
To obtain a stepsize that guarantees a uniform decrease in the cost functional we use the Armijo-rule with backtracking, \cite{Ber}.
In the sequel we denote by $\tilde{J}(\eps):=J(\psi(\eps),\eps)$, where $\psi(\eps)$ is the unique solution of the state equation for given $\eps$. Furthermore we decompose $\eps$ into its real and imaginary parts, respectively, i.e., $\eps=\eps_r+\ii \eps_\ii$.
The whole optimal control algorithm is then specified as follows:

\subsection*{OPC-algorithm}

\begin{enumerate}
\item Initialize $\eps^{old}$, $0 < c \ll 1$, $\nu \ge 1$, and $\beta > 0$.
\item \label{step2} 
\begin{enumerate}
  \item Solve the state equation $\ii \dot \psi=(H_0+H_{\eps^{old}})\, \psi$ with
  $\psi(0) = \psi_0$ (marching forward); obtain $\psi^{new}$.
  \item Solve the adjoint equation \\ 
  $\ii \dot p =(H_0^* + H_{\eps^{old}}) \, p  - q$
  with $\ii p(T) = \psi(T) - \psi_d$ (marching backwards): obtain $p^{new}$.
  \item \label{searchdirection} Determine a search direction 
  \begin{displaymath}
  \left(\begin{array}{c}d_r\\d_\ii\end{array}\right) = - G^{-1}  \left(\begin{array}{c} \nabla _r\tilde{J}(\eps^{old})\\ \nabla _\ii\tilde{J}(\eps^{old}) \end{array}\right),
\end{displaymath}
where 
\begin{align*}
&\nabla_r \tilde{J}(\eps^{old}) = \eps_r^{old} - \Re e [ p \cdot (\frac{\partial H}{\partial \eps_r}\psi)^*]^{new},\\
&\nabla_\ii \tilde{J}(\eps^{old}) = \eps_\ii^{old} - \Re e [ p \cdot (\frac{\partial H}{\partial \eps_i} \psi)^*]^{new}
\end{align*}
are the gradients for the real and imaginary parts of $\eps$, respectively, and
$G$ is a positive definite matrix. 
\end{enumerate}
\item Determine a step size $t$ such that 
\begin{equation}\label{Armijo}
\tilde{J}(\eps^{old}+t(d_r+\ii d_\ii)) < \tilde{J}(\eps^{old}) + ct \left(\begin{array}{c} \nabla _r \tilde{J}(\eps^{old})\\ \nabla _\ii\tilde{J}(\eps^{old}) \end{array}\right) \cdot \left( \begin{array}{c}d_r\\d_\ii\end{array}\right) 
\end{equation}
holds:
\begin{enumerate}
\item \label{step3a} If $t = \beta$ fulfills \eqref{Armijo}, set $\beta:=\nu \beta$, $\eps^{new}:= \eps^{old} + t (d_r+\ii d_\ii)$, and goto \ref{step2}, else
\item $\beta:=\beta/2$, goto \ref{step3a}. 
\end{enumerate} 
\end{enumerate}

Taking in step \ref{searchdirection} the matrix $G$ equal to the identity matrix leads to the usual gradient method, which can happen to converge slowly. Another idea is to use for $G$ an approximation to the Hesse matrix of $J$ with respect to the real and imaginary parts of $\eps$, which leads to quasi-Newton methods \cite{Ber}.
  
To determine the evolution of the state variable and of the adjoint variable 
we consider the implicit second-order Crank-Nicholson scheme. 
The advantage of the Crank-Nicholson scheme is that it is unconditionally 
stable and it preserves the probability density $|\psi|^2$ in case of a coherent time
evolution \cite{pdr}. For completeness, we give a brief description of the 
method. Consider the Schr\"odinger equation \eqref{schroe1}. Given the numerical solution 
at time step $m$, the value of the wave function at the next time step, $m+1$, 
is obtained solving the following problem for $\psi^{m+1}$
$$
\ii \frac{\psi^{m+1} - \psi^{m}}{\delta t}
= \frac{1}{2} H^{m+1} \psi^{m+1}  + \frac{1}{2} H^{m} \psi^{m} .
$$
Thus $\psi^{m+1}$ is given by
$$
\psi^{m+1} = (I + \ii \frac{\delta t}{2} H^{m+1})^{-1}
\, (I - \ii \frac{\delta t}{2} H^{m}) \psi^m ,
$$
where $I$ is the identity in $\C^n$. Notice the dependence $H^{m}$ from time step due to 
the presence of the control in $H$. The operator $(I + \ii \frac{\delta t}{2} H^{m+1})$ 
is a $n \times n$ complex matrix which is easily invertible 
(it can be computed analytically for small values of $n$). In case of the adjoint equation marching backwards in time the formulae above hold by inverting the time direction.

Accuracy of the solution obtained by integrating in time using the Crank-Nicholson scheme 
(or $\theta=1/2$-method) is known, and we therefore report only the main result. Denote with 
$e^m=\psi(t_m)-\psi^m$, $m=1,\dots,N_{steps}$, the error at each time step between the 
continuous solution $\psi(t)$ and its numerical approximation $\psi^m$. Then, assuming 
$e^0=0$, there exist constants $L$ and $C$ such that
$$
|e^m| \le C \, {\delta t} ^2 \left[ \exp \left( L \frac{t_m}{1-L \delta t /2} \right)  -1 \right] ,
$$
where $L$ is a Lipschitz constant and $C$ is proportional to $\sup_{[0,T]} |\psi^{'''}|$.

\section{Model system}\label{sec:model}

\begin{figure}
\[\begin{xy}0;<12mm,0mm>:
  (0,-0.3);(1,-0.3)*{}**@{-},   
  (2,0);(3,0)*{}**@{-},
  (1,2);(2,2)*{}**@{-},  
  (0.5,-0.6)*{\psi_1},
  (2.5,-0.3)*{\psi_2},
  (1.5, 2.3)*{\psi_3},
  (3.1,-0.15)*{\delta},
  (3,-0.3);(2.9,-0.3)*{}**@{-},
  \ar@{<=>}(0.5,-0.2);(1.2,1.9)
  \ar@{<~} (1.5,0);(1.5,1.9)
  \ar@{<=>}(2.5,0.1);(1.8,1.9)
  \ar@{.}(2.95,-0.28);(2.95,0)
\end{xy}\]
\caption{Prototypical $\Lambda$-type level scheme used in our calculations: 
$\phi_1$ and $\phi_2$ are long-lived states, whereas $\phi_3$ is a short-lived state which is optically coupled to both $\phi_1$ and $\phi_2$ (for details see text); wiggled line indicates relaxation and dephasing of state $\phi_3$.}
\label{fig:scheme}
\end{figure}
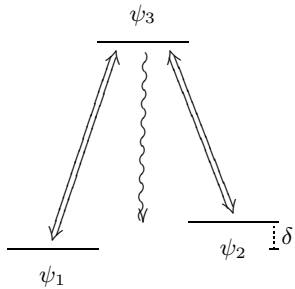

To demonstrate the applicability of our approach, as a first representative example in this paper we consider the three-level system depicted in Fig.~\ref{fig:scheme}, which consists of: two long-lived states $\phi_1$ and $\phi_2$, which are energetically separated by some amount $\delta$; a state $\phi_3$, which has a finite lifetime because of environment coupling (wiggled line). Such $\Lambda$-type configurations have a long-standing history in quantum optics and have been demonstrated successful in the explanation of many coherence-phenomena in atomic systems \cite{scully:97,mandel:95,bergmann:98}; more recently, similar configurations have received increasing interest also in semiconductor quantum dots \cite{hohenester.apl:00,hohenester:02,brandes:00}.

Within this scheme, the system's time evolution is governed by the effective Hamiltonian \cite{bergmann:98,hohenester:02}:

\begin{equation}
H_0 = \frac 1 2
\left(\begin{array}{ccc}
     -\delta    &       0           &   0  \\
     0          &    \delta         &   0  \\
     0          &       0           &  -\ii \gamma_o \\
\end{array}\right) ,
\label{h0}
\end{equation}

\noindent where the term $-\ii\gamma_o$ accounts for environment losses (e.g., spontaneous photon emissions). Furthermore, the coupling to the external field reads:

\begin{equation}
H_\eps = -\frac{1}{2}
\left(\begin{array}{ccc}
     0          &       0           &   \mu_1 \eps  \\
     0          &       0           &   \mu_2 \eps  \\
\mu_1 \eps^*  &  \mu_2 \eps^*   &   0
\end{array}\right),
\label{he}
\end{equation}

\noindent where $\mu_1$ and $\mu_2$ describe the coupling strengths of states $\phi_1$ and $\phi_2$ to the inter-connecting state $\phi_3$ (e.g., optical dipole matrix elements); note that in Eqs.~\eqref{h0} and \eqref{he} we have implicitly assumed the usual rotating-wave approximation \cite{scully:97,hohenester:02}. Initial and final states are then given by:
$$
\psi_0 = \left(\begin{array}{c} 1 \\ 0 \\ 0 \end{array}\right), \qquad
\psi_d = \left(\begin{array}{c} 0 \\ e^{-\ii \delta T} \\ 0 \end{array}\right) ,
$$

\noindent and for the optimality equations (\ref{staeq}) and (\ref{adjeq}) we obtain
$$
\eps  =  -     \frac{1}{2\gamma} \Re e [ p \cdot (H_1\psi)^*]
         - \ii \, \frac{1}{2\gamma} \Re e [ p \cdot (H_2\psi)^*] ,
$$
with
\begin{equation*}
H_1 =
\left(\begin{array}{ccc}
     0          &       0           &   \mu_1   \\
     0          &       0           &   \mu_2   \\
\mu_1           &  \mu_2            &   0
\end{array}\right) , \qquad
H_2 =
\left(\begin{array}{ccc}
     0          &       0           &   \ii \mu_1   \\
     0          &       0           &   \ii \mu_2   \\
 -\ii \mu_1     &  -\ii \mu_2     &   0
\end{array}\right) .
\label{hc}
\end{equation*}

\section{Results}\label{sec:results}

Assuming that the system is initially prepared in state $\phi_1$, in the following we address the question: what is the most efficient way to bring the system from $\phi_1$ to $\phi_2$? Since the direct optical transition between $\phi_1$ and $\phi_2$ is assumed to be forbidden we have to use $\phi_3$ as an auxiliary state; however, intermediate population of $\phi_3$ introduces losses through environment coupling. Thus, which sequence of laser pulses minimizes the population of level $\phi_3$? 

\subsection{Simplified model system}

To gain insight into the general trends, in the following we shall discuss a somewhat simplified model system (results of our complete calculations will be presented further below). We assume that the three-level system of Fig.~\ref{fig:scheme} is subject to two fields $\eps_1(t)=g(t-\frac {t_o} 2)\exp\ii\frac \delta 2 t$ and $\eps_2(t)=g(t+\frac {t_o} 2)\exp-\ii\frac \delta 2 t$, respectively, where $g(t)$ denotes a Gaussian envelope with full-width of half maximum $\tau$; in addition, we assume that the first pulse $\epsilon_1$ (centered at time $t_o/2$) only affects the 1--3 transition, and the second pulse $\epsilon_2$ (centered at time $-t_o/2$) only the 2--3 one; such approximation corresponds to the case that $\delta$ is much larger than $\mu_1\epsilon$ and $\mu_2\epsilon$ \cite{scully:97,bergmann:98}.

\begin{figure}
\includegraphics[width=0.75\columnwidth]{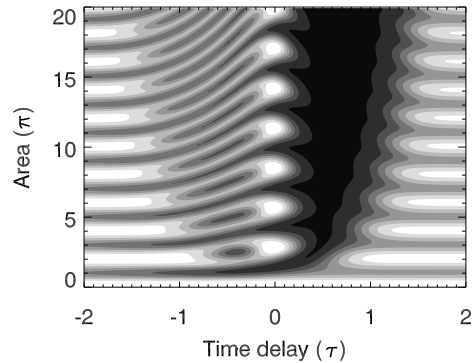}
\caption{
Results of our calculations for the simplified model system described in the text, and using two laser pulses with Gaussian envelopes (with full-width of half maximum $\tau$). Before the pulse sequence the system is in state $\phi_1$; black (white) areas correspond to the situation that after the pulse sequence the population of $\phi_2$  is one (zero). Negative (positive) time delays correspond to the situation that the $\eps_1$ pulse excites the system before (after) the $\eps_2$ one, and the pulse area is defined as $\int_{-\infty}^\infty dt\;g(t)$ (we use $\mu_1=\mu_2=1$ and $\gamma_o=0.2\tau^{-1}$).
}\label{fig:stirap}
\end{figure}

In Fig.~\ref{fig:stirap} we present results for this simplified model system for different time delays $t_o$ between the two pulses and for different pulse areas (as defined in the figure caption). As regarding the general trends, we observe from Fig.~\ref{fig:stirap} that successful population transfer between states $\phi_1$ and $\phi_2$ can be achieved for both negative and positive time delays $t_o$. In the first case, the pulse $\eps_1$ excites the system {\em before}\/ the $\eps_2$ one; here, $\eps_1$ brings the system from $\phi_1$ to the auxiliary state $\phi_3$, and $\eps_2$ channels the population between $\phi_3$ and the final state $\phi_2$; apparently, the efficiency of this transfer, which is known as the stimulated emission pumping technique \cite{bergmann:98}, becomes maximal when the pulse areas are odd multiples of $\pi$. In contrast, for positive time delays, i.e., when the $\eps_1$ pulse excites the system {\em after}\/ the $\eps_2$ one, the population transfer is not achieved through intermediate shelving of population; for that reason, the sequence of laser pulses is called {\em counter-intuitive},\/ and the whole process has been given the name stimulated Raman adiabatic passage (STIRAP) \cite{bergmann:98}. This STIRAP process exploits the renormalizations of quantum states in presence of the strong laser fields, and population transfer is achieved by keeping $\psi_3$ negligible throughout; see Ref.~\cite{bergmann:98} for an excellent review.

\subsection{Optimal control}

\begin{figure}
\includegraphics[width=0.48\columnwidth]{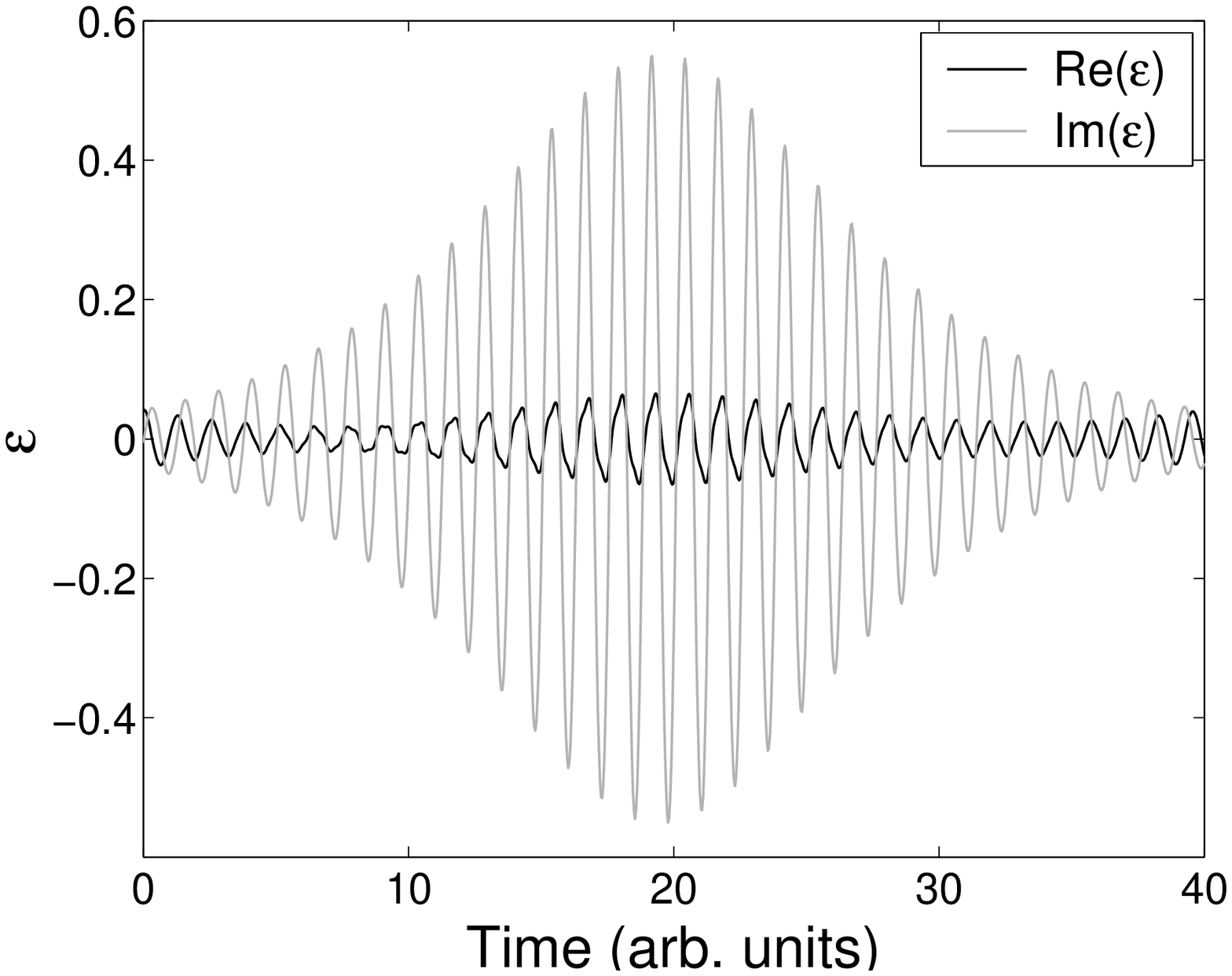}
\includegraphics[width=0.48\columnwidth]{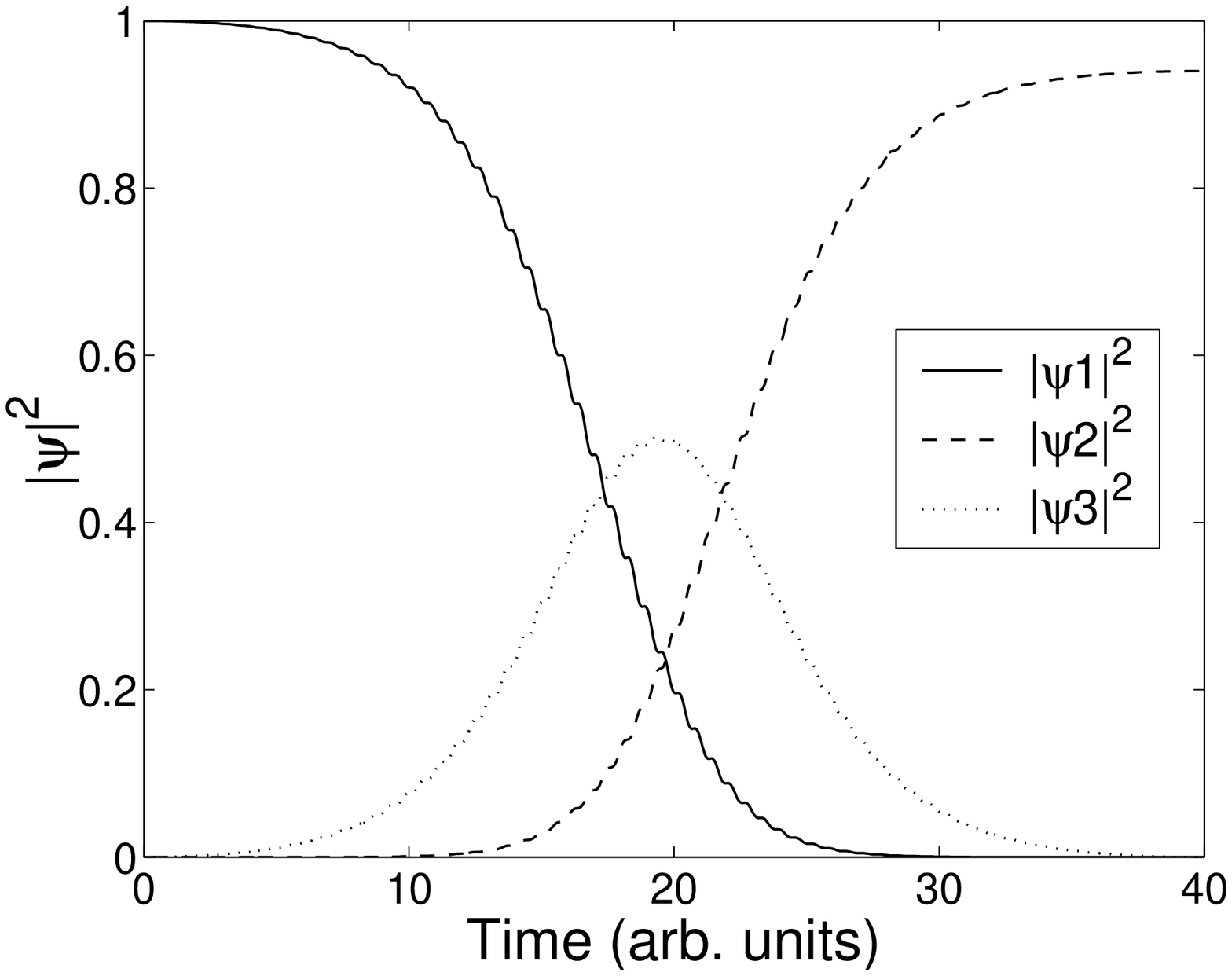}
\caption{
Results of our optimal-control calculations for $\gamma=0.001$, $\alpha=0$: Optimal control (left) and transients of $\psi_1$, $\psi_2$, $ \psi_3$ (right).}\label{fig:intuitive}
\end{figure}

We next consider the population transfer for our complete model system of Eqs.~\eqref{h0} and \eqref{he}, i.e., $\eps$ affects both the 1--3 and 2--3 transitions, within the framework of {\em optimal control}\/ (see Ref.~\cite{boscain:02} for a related optimization analysis of the coherent dynamics). As will become apparent from our following discussion, we can change the characteristics of the transfer process between the two limiting cases through different penalization of $\psi_3$, i.e., through modification of the weight $\alpha=\alpha_3$ in the cost functional \eqref{minJ}. In the solution of Eqs.~(\ref{staeq}--c) we use $c=0.001$, $\nu=1.1$, and initialize $\beta$ with $0.2$ (we checked that our results do not depend decisively on these parameters). Furthermore, we use $\mu_1=\mu_2=1$, $\delta=10$, $\gamma_o=0.01$, and consider a final time $T=40$. To update the matrix $G$ in our algorithm we use the Broyden--Fletcher--Goldfarb--Shanno (BFGS) formula \cite{Ber}. Unless otherwise specified, $\eps=0.1 \exp(\ii\delta t)$ for the initialization of the algorithm, and we stop the iteration if the norm of the gradient is less then $10^{-5}$.

Figure \ref{fig:intuitive} shows results for (a) the control field and (b) the quantum-state population for $\alpha=0$ (i.e., no penalization of $\psi_3$). We observe that the population is channeled through occupation of the interconnecting $\phi_3$ state, in close resemblance to the stimulated emission pumping process; indeed, analyzing the Fourier transform of the control field, Fig.~\ref{fig:intuitive}a, we find two strong contributions at frequencies $-\delta/2$ and $\delta/2$, where the first one dominates at times below $20$ and the latter one in the second half of the transfer process.

\begin{table}[b]
\caption{Results of our calculations for different values of $\gamma$.}
\begin{ruledtabular}
\begin{tabular}{ccccc}
 $\gamma$ & $|\psi(T) - \psi_d |_{\C^3}^2$ & $J$      & $iter$ 
\\ \hline
 $10^{-1}$         & $5.76 \cdot 10^{-3}$          & $3.93 \cdot 10^{-2}$   & $26$     \\ 
 $10^{-2}$         & $2.68 \cdot 10^{-3}$          & $5.53 \cdot 10^{-3}$   & $63$     \\ 
 $10^{-3}$         & $7.44\cdot 10^{-4} $          & $1.11\cdot 10^{-3} $   & $123 $  \\
 $10^{-4}$         & $1.68\cdot 10^{-4} $          & $2.44 \cdot 10^{-4}$   & $360 $   \\
\end{tabular}
\end{ruledtabular}
\end{table}

Further insight into the pertinent calculation parameters can be obtained from table I which reports the influence of $\gamma$, Eq.~\eqref{minJ}, on: the tracking error $|\psi(T)-\psi_d |_{\C^3}^2$, i.e., the measure of how accurately the final state is reached; the value of the cost functional $J$; and the number $iter$ of required minimization steps. Quite generally, we observe that allowing stronger field strengths through choice of smaller $\gamma$-values results in a more efficient population transfer, as one would already expect from the discussion of the simplified model system: here, the increased control field allows for faster $1\to 3$ and $3\to 2$ transitions and, in turn, smaller environment losses due to population of $\phi_3$. We also verified that the on- and off-switching of the control fields can be controlled by replacing $\eps$ in the state equation by $\xi(t)\eps$, where $\xi(t)$ is a function that smoothly approaches zero at early and late times; such additional control might be required to account for the limited laser resources in experiment. We finally note that  the number of required iterations of the algorithm significantly increases as $\gamma$ decreases, which we attribute to the increasing singularity of the optimal-control problem as $\gamma\rightarrow 0^+$.

\begin{figure}
\includegraphics[width=0.48\columnwidth]{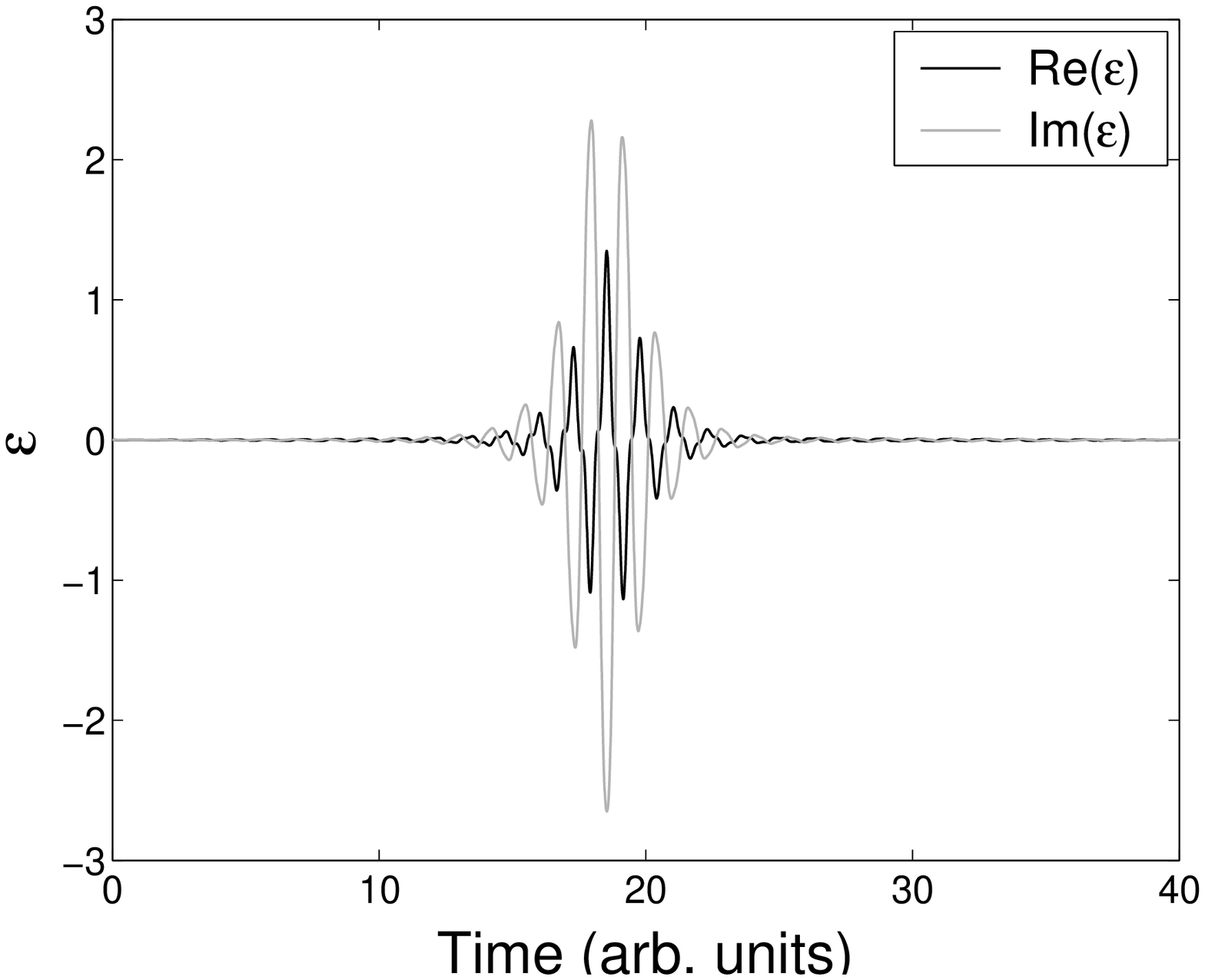}
\includegraphics[width=0.48\columnwidth]{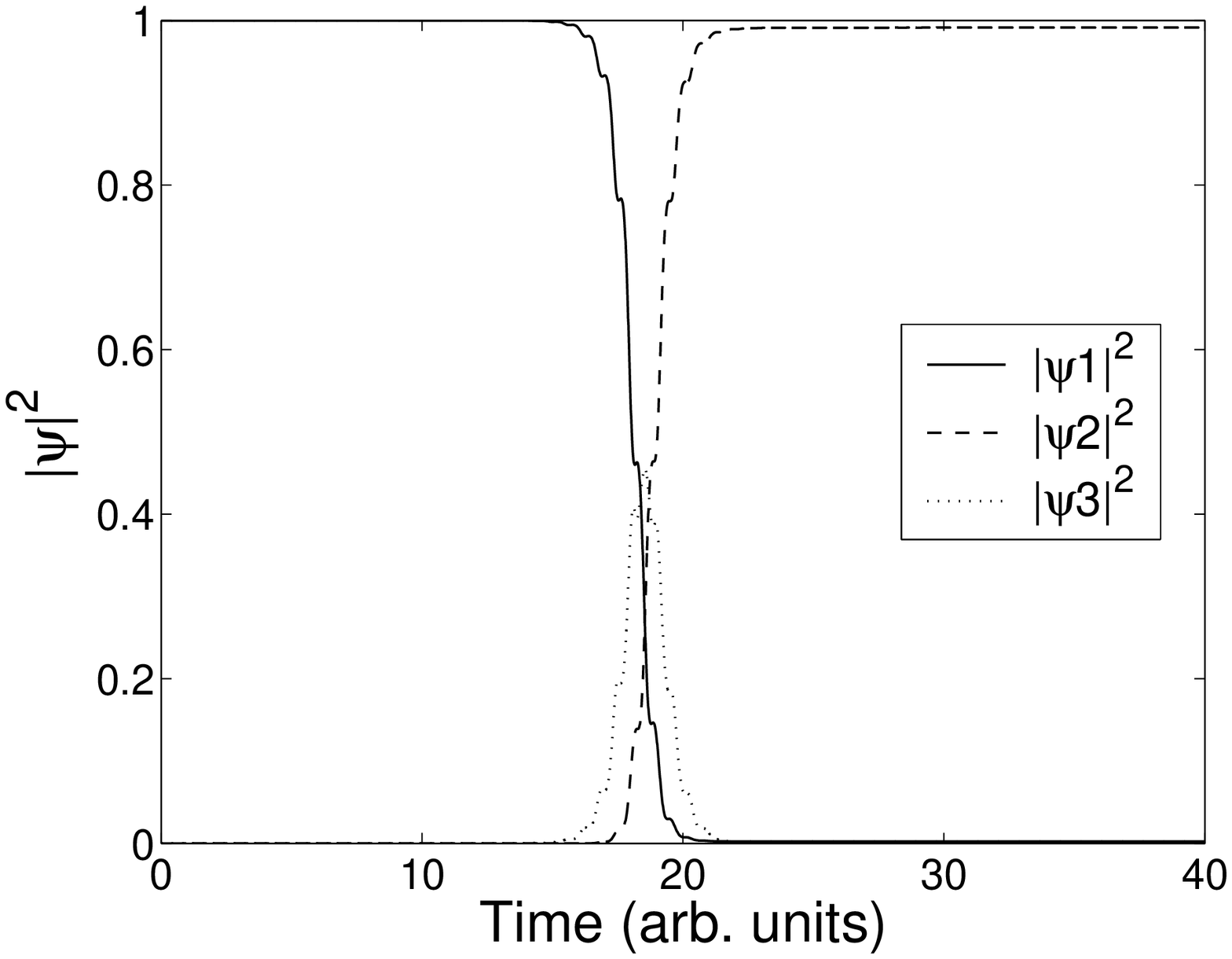}
\caption{
Same as Fig.~\ref{fig:intuitive} but for $\alpha=0.01$.
}\label{fig:counter-intuitive}
\end{figure}

Finally, in Fig.~\ref{fig:counter-intuitive} we show the influence of the penalization of the $\phi_3$-population through finite values of $\alpha_3$, Eq.~\eqref{minJ}, on the solution of the optimal-control problem, which results in a strong reduction of the population-transfer time. From the Fourier transform of the control field of Fig.~\ref{fig:counter-intuitive} we furthermore infer that here the time ordering of the two dominant frequency components is reversed as compared to the $\alpha=0$ case, thus making this excitation scenario similar to the STIRAP process. 

\ \\ \vspace*{-0.5cm}

\section{Conclusions}\label{sec:conclusions}

In conclusion, we have presented a theoretical analysis of optimal control of quantum nanostructures. A general solution scheme for the optimization of an external control (e.g., lasers pulses) has been developed, which allows to channel the system's wavefunction between two given states in its most efficient way; physically motivated constraints, such as limited laser resources or population suppression of certain states, can be accounted for through a general cost functional. A computer algorithm for the solution of the optimal control problem has been derived and analyzed in detail. Finally, we have demonstrated the applicability of our approach for a generic three-level quantum system, and we have identified the pertinent calculation and convergence parameters.

Apparently, the true strength of optimal control can only be appreciated in the study of higher-dimensional systems where the control strategies can no longer be grasped from simple considerations, which are needed, e.g., for the design of quantum gates in future quantum registers. There, the high flexibility of our present approach, which solely relies on the state equation and a general functional accounting for the control constraints, renders {\em optimal control}\/ as an ideal tool for both theoretical modeling as well as experimental support. In this respect, it will be necessary to develop more efficient numerical methods for the solution of the {\em bilinear}\/ optimal control problem of our present concern. Future work will also address applications beyond the presently studied three-level scheme.

\acknowledgments
Work supported in part by the {\em Fonds zur F\"orderung der wissenschaftlichen Forschung}\/ (FWF) under SRC 03 ''Optimization and Control'' and P15752.

\begin{appendix}

\section{}

In this section we present the proof of Theorem 1.

\begin{proof}{\em Step 1:} We first calculate an a-priori estimate for the solution of the state 
equation \eqref{schroe1}. For given $\eps \in L^2(\C, [0,T])$ the equation \eqref{schroe1} is a linear ordinary differential equation, which therefore has a unique solution $\psi\in H^1(\C^n,[0,T])$. We 
now write Eq.~\eqref{schroe1} in integral form:
\begin{equation}
\label{intform}
\ii \psi(t) = \psi_0 + \int_0^t(H_0+H_\eps)\psi(s)\,ds \:\mbox{ for }\: 0\le t\le T.
\end{equation}
Taking the Euclidean norm in $\C^n$ on both sides of \eqref{intform} and using the triangle 
inequality on the right-hand side results in
\begin{equation*}
||\psi(t)|| \le  ||\psi_0|| + \int_0^t(||H_0||+||H_\eps||)||\psi(s)||\,ds,
\end{equation*}
where for the matrices $H_0$ and $H_\eps$ $||\cdot ||$ denotes the induced matrix norm. 
We can now apply Gronwall's inequality \cite{Fat}, and obtain
\begin{equation*}
||\psi(t)|| \le  ||\psi_0|| \exp \left( \int_0^t(||H_0||+||H_\eps||)\,ds\right) \mbox{ for } 0\le
t\le T.
\end{equation*}
Using equation \eqref{est H_eps} and squaring and integrating the above inequality results in 
\begin{equation*}
||\psi(t)||_{L^2(\C^n,[0,T])} \le  ||\psi_0|| K_1 \exp \left( K_2 +K_3||\eps||_{L^2(\C,[0,T])}\right)
\end{equation*}
with constants $K_1,K_2,K_3>0$ which do not depend on $\eps$. Furthermore, using the state 
equation \eqref{schroe1} yields that for all $\eps\in L^2(\C,[0,T])$ with $||\eps||_{L^2(\C,[0,T])}
\le K$ the corresponding states $\psi(\eps)$ are bounded in $H^1$, 
i.e., $||\psi(\eps)||_{H^1(\C^n,[0,T])}\le \bar{K}$ for some $\bar{K}>0$.\\

{\em Step 2:} Let $(\eps_k)_{k\ge 1}$ be a minimizing sequence for $J$, i.e.
\begin{equation*}
\lim_{k\to \infty} J(\psi_k,\eps_k) = \inf_{(\psi,u)\: \mbox{\scriptsize satisfies }
\eqref{schroe1}} J(\psi,\eps),
\end{equation*}
where we denote by $\psi_k=\psi(\eps_k)$ the unique solution of \eqref{schroe1} for 
given $\eps_k$. Due to the fact that $J(\psi(\eps),\eps) \rightarrow \infty$ as $||\eps||_{L^2(\C,[0,T])}\rightarrow \infty$, we get that the sequence $(\eps_k)_{k\ge 1}$ 
is bounded in $L^2(\C,[0,T])$. Since the unit ball in a Hilbert space is weakly compact, 
there exists a weakly to an $\bar{\eps}\in L^2(\C,[0,T])$ convergent subsequence, which 
we again denote by $(\eps_k)_{k\ge 1}$. Step 1 above ensures that the corresponding 
sequence $(\psi_k)_{k\ge 1}$ is bounded in $H^1(\C^n,[0,T])$; thus, again by 
choosing a proper subsequence
\begin{equation*}
\psi_k\rightharpoonup \bar{\psi} \:\:\mbox{weakly in } H^1(\C^n,[0,T]),
\end{equation*}
it follows from the Sobolev imbedding theorem \cite{Ada} that 
\begin{equation*}
\psi_k\rightarrow \bar{\psi} \:\:\mbox{strongly in } L^2(\C^n,[0,T]) \mbox{ and in } C^0(\C^n,[0,T]).
\end{equation*}

We can now show that $(\bar{\psi},\bar{\eps})$ is a solution of the optimal control problem. 
From the definition of $\psi_k,\eps_k$ we have
\begin{equation}\label{kintform}
\ii \psi_k(t) = \psi_0 + \int_0^t(H_0+H_{\eps_k})\psi_k(s)\,ds\:\mbox{  for }\: 0\le t\le T.
\end{equation}
Next, we consider the limit in \eqref{kintform}. The weak $L^2$-convergence of $\eps_k$ to 
$\bar{\eps}$ implies weak convergence also for the complex conjugates, i.e.~$\eps_k^* 
\rightharpoonup \bar{\eps}^*$ in $L^2(\C,[0,T])$. Strong convergence of $\psi_k$ to $\psi$ in 
$L^2(\C,[0,T])$ allows to go to the limit as $k\to \infty$ on the right-hand side of \eqref{kintform}. 
Thus we find
\begin{equation*}
\ii \bar{\psi}(t) = \psi_0 + \int_0^t(H_0+H_{\bar{\eps}})\bar{\psi}(s)\,ds,
\end{equation*}
which shows that $\bar{\psi}=\psi(\bar{\eps})$, or equivalently that $(\bar{\psi},\bar{\eps})$ 
fulfills \eqref{schroe1}. We finally obtain

\begin{widetext}

\begin{align*}
J(\bar{\psi},\bar{\eps}) = &  \frac{1}{2} | \bar{\psi}(T) - \psi_d |_{\C^n}^2 + \frac{\gamma}{2} ||
\bar{\eps} ||_{L^2(\C,[0,T])}^2 + \frac{\alpha}{2} \sum_{j=1}^n || \bar{\psi}_j ||_{L^2(\C,[0,T])}^2 \\[0.2cm]
 \le &  \frac{1}{2} \lim_{k\to\infty}| \psi_k(T) - \psi_d |_{\C^n}^2
+ \frac{\gamma}{2}\liminf_{k\to \infty} || \eps_k ||_{L^2(\C,[0,T])}^2 + \frac{1}{2}
\lim_{k\to \infty} \sum_{j=1}^n\alpha_j || \psi_{k,j} ||_{L^2(\C,[0,T])}^2\\[0.2cm] = & \inf_{(\psi,\eps)\:
\mbox{\scriptsize satisfies } \eqref{schroe1}} J(\psi, \eps),
\end{align*}

\end{widetext}
where we used the lower-semicontinuity of the $L^2$-norm. Thus we have proved that $(\bar{\psi},\bar{\eps})$ is a solution to problem of Eqs.~\eqref{schroe1} and \eqref{minJ}.
\end{proof}

\end{appendix}

\end{document}